\documentclass[conference,10pt]{IEEEtran}
\usepackage{graphicx}
\usepackage{acronym}
\usepackage{amsmath}
\usepackage{balance}

\usepackage[ruled,linesnumbered,norelsize]{algorithm2e}
\SetAlFnt{\footnotesize}

\usepackage{cite}
\usepackage{doi}

    \providecommand{\matr}[1]{\begin{bmatrix} #1 \end{bmatrix}}
    
    \providecommand{\smatr}[1]{\left[\begin{smallmatrix} #1 \end{smallmatrix}\right]}
    
    \DeclareMathOperator{\N}{N}                                 % normal distribution
         
	\newcommand{\noise}{\varepsilon}

	\newcommand{\Enoise}{\noise_{\mathrm{E}}}
	\newcommand{\Enoisei}[1]{\noise_{\mathrm{E},#1}}

	\newcommand{\PE}{P_{\mathrm{E}}}
	
	\newcommand{\Nnoise}{\noise_{\N}}
	\newcommand{\measfun}{h}
	\newcommand{\transloc}{x_t}
	\newcommand{\recloc}{x_r}
	\newcommand{\nnoise}{n}	
	\newcommand{\transfun}[1]{f\left(#1\right)}
		
		\newcommand{\pchipfun}[1]{f_\text{P}\left(#1\right)}
		
\newcommand{\nonlcov}{\Omega}

\newcommand{\constant}{b}
\newcommand{\Pxy}{{P_{x\measfun(x)}}}
\newcommand{\Pyx}{{P_{\measfun(x)x}}}

\newcommand{\Pzz}{{P_{zz}}}
\newcommand{\hatP}{{}\hat{P}}
\newcommand{\hatPzz}{{{ \hatP_{zz}}}}

\newcommand{\Pzy}{{P_{z\measfun(z)}}}
\newcommand{\Pyz}{{P_{\measfun(z)z}}}

\newcommand{\Pxx}{{P_{xx}}}
\newcommand{\Pyy}{{P_{\measfun(x)\measfun(x)}}}

\newcommand{\Pee}{{P_{\noise\noise}}}
\newcommand{\Pex}{{P_{\noise x}}}
\newcommand{\Pxe}{{P_{x \noise }}}
\newcommand{\polypoints}{m}
\newcommand{\nonlnoise}{{\varepsilon_\nonlcov}}
\newcommand{\noisecov}{R} % Measurement noise covariance 		

\newcommand{\mean}{\mu}
\newcommand{\state}{x}
\newcommand{\measnoise}{\varepsilon}
\newcommand{\cov}{P}
\newcommand{\statecov}{{\cov_{\state\state}}}

\newcommand{\knotpos}{s}
\newcommand{\sample}{\knotpos}

\newcommand{\knotval}{y}

\newcommand{\kalmangain}{K}
\newcommand{\statemean}{{\mean_{\state}}}
\newcommand{\noisemean}{{\mean_{\noise}}}

\newcommand{\priormean}{\statemean^-}
\newcommand{\priorcov}{\statecov^-}

\newcommand{\augstate}{z}
\newcommand{\augcov}{{\cov_{\augstate\augstate}}}
\newcommand{\augmean}{{\mean_\augstate}}
\newcommand{\meas}{y}

\newcommand{\predmeas}{\hat{\meas}}

\newcommand{\jacobian}{J}

\newcommand{\statedim}{d}
\newcommand{\derivative}{d}
\newcommand{\slope}{\Delta}
\newcommand{\measdim}{m}

\newcommand{\iterations}{N}

\renewcommand{\d}{\mathrm{d}}
\newcommand{\diag}[1]{\mathrm{diag}\left( #1 \right)}

\newcommand{\innocov}{S}

\newcommand{\pitch}{\theta}
\newcommand{\yaw}{\psi}
\newcommand{\roll}{\phi}

\newcommand{\gain}{K}
\title{Kalman filtering with empirical noise models}
\author{\IEEEauthorblockN{Matti Raitoharju\IEEEauthorrefmark{1}, Henri Nurminen \IEEEauthorrefmark{2}, Demet Cilden-Guler \IEEEauthorrefmark{3}, and Simo S\"arkk\"a\IEEEauthorrefmark{1}}\\
 \IEEEauthorblockA{\IEEEauthorrefmark{1}Department of Electrical Engineering and Automation, Aalto University, Espoo, Finland\\
}
\IEEEauthorblockA{\IEEEauthorrefmark{2}Here, Tampere, Finland}
\IEEEauthorblockA{\IEEEauthorrefmark{3}Faculty of Aeronautics and Astronautics, Istanbul Technical University, Turkey
}}
\IEEEoverridecommandlockouts
\IEEEpubid{\parbox{\columnwidth}{This is an author accepted version of a paper published in conference proceedings of the ICL-GNSS 2021 \hfill \copyright2021 IEEE} \hspace{\columnsep}\makebox[\columnwidth]{ }}

\begin{document}
\newacro{PLF}{Posterior Linearization Filter}
\newacro{RUF}{Recursive Update Filter}

\newacro{RMS}{Root Mean Square}
\newacro{SLR}{statistical linear regression}
\newacro{KF}{Kalman Filter}
\newacro{KFE}{Kalman Filter Extension}

\newacro{EKF}{Extended Kalman Filter}
\newacro{UKF}{Unscented Kalman Filter}
\newacro{CKF}{Cubature Kalman Filter}

\newacro{IPLF}{Iterated Posterior Linearization Filter}
\newacro{KLD}{Kullback-Leibler Divergence}
\newacro{GGF}{General Gaussian Filter}

\newacro{GRUF}{Generalized Recursive Update Filter}
\newacro{RORF}{Recursive Outlier-Robust Filter}

\newacro{IGRF}{International Geomagnetic Reference Field}

\newacro{KDE}{Kernel Density Estimation}
\newacro{DPLF}{Damped Posterior Linearization Filter}
\newacro{PCHIP}{Piecewise Cubic Hermite Interpolating Polynomial}
\newacro{NLOS}{Non-Line-of-Sight}
\newacro{UWB}{Ultra Wideband}
\newacro{cdf}{cumulative distribution function}
\newacro{pdf}{probability density function}

\maketitle
\IEEEpubidadjcol
\begin{abstract}
Most Kalman filter extensions assume Gaussian noise and when the noise is non-Gaussian, usually other types of filters are used. These filters, such as particle filter variants, are computationally more demanding than Kalman type filters. In this paper, we present an algorithm for building models and using them with a Kalman type filter when there is empirically measured data of the measurement errors. The paper evaluates the proposed algorithm in three examples. The first example uses simulated Student-$t$ distributed measurement errors and the proposed algorithm is compared with algorithms designed specifically for Student-$t$ distribution. Last two examples use real measured errors, one with real data from an \ac{UWB} ranging system, and the other using low-Earth orbiting satellite magnetometer measurements. The results show that the proposed algorithm is more accurate than algorithms that use Gaussian assumptions and has similar accuracy to algorithms that are specifically designed for a certain probability distribution. 
\end{abstract}
\acresetall
\section{Introduction}
A measurement model defines a relationship between a state and the value of the measurement. In practice, all measurements contain uncertainty or noise and this uncertainty can be taken into account in the measurement model. Often the noise is modeled with normal distributions. The use of normal distributions can be justified with the central limit theorem, which states that the sum of independent random variables tends towards a normal distribution.  Often the main reason for the use of normal distributions is its convenience; many algorithms are based on the normality assumption and normal distributions have only two parameters, mean and (co)variance. However, in reality the Gaussian assumption may not hold. Measurement data may, for example, contain outliers \cite{outliers} or be skewed \cite{henriletter}.

Bayesian filtering is a framework for estimating the state given noisy measurements and a state model that describes how the state evolves in time. Closed-form solutions for the Bayes filter exist only in some special situations, e.g.\, \ac{KF} is the Bayes filter when everything is linear and Gaussian. Various~\acp{KFE} have been developed for estimation with nonlinear state or measurement models. For measurements with outliers there are algorithms called robust \acp{KF} \cite{2015arXiv150904072W,agamennoni}. There are also variational Bayes based algorithms for non-Gaussian noises \cite{piche2012d} for Student-$t$ noise and  \cite{henriletter} for skew-$t$ distribution.

In \cite{GRUF} and \cite{8260875} an iterative \ac{KFE} that can do state estimation when the measurement model noise is non-Gaussian were presented. Main difference between algorithms in \cite{GRUF} and \cite{8260875} is that the former is based on modeling the noise distribution using a nonlinear transformation of Gaussian and latter uses computation of moments of different distributions.

Particle filters can also be used for estimation with non-Gaussian noise. Particle filters are often computationally more demanding than \acp{KFE}, especially when the dimension of the state is large.

In modeling non-Gaussian noise there is the problem of selecting proper distribution for the measurement noise. One can seek for a named distribution, such as Student-$t$, matches with the noise distribution, or one can use approximate methods, such as \ac{KDE}. Also one has to keep in mind when selecting the noise model whether a suitable estimation algorithm is available for that specific noise model.

In this paper, we propose an approximative way to model noise distribution using samples from the noise distribution and a spline.  We also develop an iterative \ac{KFE} that can use the approximate noise model for state estimation and show that it can provide accurate results. The proposed algorithm uses only a single Gaussian for the state estimate. This reduces the complexity of algorithm compared to filters that use Gaussian mixtures to model the measurement \ac{pdf}, such as \cite{RAITOHARJU2020107330}, where the number of Gaussians increase as each measurement arrives and then component reductions are applied to keep the computational cost feasible.

The rest of this paper is organized as follows. In Section~\ref{sec:background}, used measurement model  is presented. In Section~\ref{sec:noisetrans}, we present the approximative way to model the noise distribution. We develop the iterative \ac{KFE} in Section~\ref{sec:ite}. We give results of evaluation of the accuracy of the proposed algorithms with simulated and real data in Section~\ref{sec:results}. Section~\ref{sec:conclusions} concludes the paper.

\section{Measurement model}
\label{sec:background}

In this paper, we concentrate on a single update of a prior. The prior is modeled with a normal distribution. We use a measurement model of the form
\begin{equation}
	y =  \measfun(x) + \Enoise, \label{equ:measmodelorig}
\end{equation}
where $y$ is the measurement value, $\measfun(\cdot)$ is the measurement function,  $x$ is the state, and $\Enoise$ is measurement noise whose distribution is not necessarily Gaussian. Furthermore, if the measurement is multidimensional, we assume that the components of  $\Enoise$ are independent. In case of multidimensional measurements each measurement component is treated separately. We use the approach from \cite{GRUF} and use a mapping function 
\begin{equation}
	\Enoise=\transfun{\Nnoise}, \label{equ:tran}
\end{equation}
where $\Nnoise$ is distributed as a standard normal variable, $\transfun{\cdot}$ is a strictly increasing function, and  $\transfun{\Nnoise}$ has the same distribution as $\Enoise$. If the measurement is multidimensional, then each component of $f(\cdot)$ are increasing functions. Thus, the measurement model becomes
\begin{equation}
	y =  \measfun(x) + \transfun{\Nnoise} \label{equ:measnoise}
\end{equation}
and all random variables are normal distributed.

In \cite{GRUF}, $\transfun\cdot$ was selected to be
\begin{equation}
\transfun\Nnoise= F^{-1}(\Phi(\Nnoise)), \label{equ:Trans}
\end{equation} where $F^{-1}(\cdot)$ is the inverse of the \ac{cdf} of $\Enoise$ and $\Phi(\cdot)$ is the \ac{cdf} of a standard normal distribution. In the next section, we propose to use a strictly increasing spline to model the transformation function.

\section{Finding a transformation for the noise}
\label{sec:noisetrans}

In general, the choice of  $\transfun{\cdot}$ is not unique. We require that it is strictly increasing\footnote{the choice between strictly increasing or decreasing $\transfun{\cdot}$ is arbitrary}. The reason for this requirement is that then there is always one to one mapping in~\eqref{equ:tran}. This is essential in the filtering step where we will optimize the value of $\transfun{\Nnoise}$, which will be unique as value $\transfun{\Nnoise}$ increases as $\Nnoise$ increases compared to situation where there could be multiple local optima.

%To see why the requirement for strictly monotonic\footnote{choice of having strictly increasing or decreasing $\transfun{\cdot}$ is arbitrary}  $\transfun{\cdot}$ is necessary, we consider \ac{EKF}, which would approximate \eqref{equ:measnoise} locally as
%\begin{equation}
%	\begin{aligned}
%	& \left. \measfun(x) + \transfun{\Nnoise} \right|_{x_0, \Nnoise(0)}   \approx  \\ & \measfun'(x_0) (x-x_0) + \transfun{\Nnoise(0)}'(\Nnoise-\Nnoise(0) ).
%	\end{aligned}
%\end{equation}
%If $\transfun{\Nnoise}$ is not strictly monotonic $\transfun{\Nnoise(0)}'=0$ at some point and the measurement function would not depend on the noise term at all and the measurement would be treated as a noise free and the posterior estimate would become singular. \acp{KFE} that use \ac{SLR} do linearization in a larger area, but having a strictly increasing function produces always an increasing linearization.

Transformation using \eqref{equ:Trans} requires the knowledge of the inverse \ac{cdf} ($F^{-1}$), which we do not assume to be known. To model an empirically determined distribution, we assume that we have $\nnoise$ samples from the real non-Gaussian noise $\Enoise$.  One option for approximating \ac{cdf} of $\Enoise$ from samples is
\begin{equation}
F(s) \approx \PE(\sample) = \frac{ \lvert \left\{ \Enoisei{i} \mid \Enoisei{i} \leq \sample  \right\} \rvert}{\nnoise} \label{equ:cdfapp},
\end{equation}
where $\lvert \cdot \rvert$ is the size of the set and $\sample$ is the value where the \ac{cdf} approximation is computed. However, this approximation is not suitable for our purposes as such, because it is not invertible and the function is not continuous as there are discrete steps.

We can use \eqref{equ:cdfapp} for choosing points $\sample_i$ that are  such that 
\begin{equation}
	\PE(\sample_i)<\PE(\sample_{i+1}).
\end{equation}
To avoid approximating directly the inverse of \ac{cdf}, we substitute $\Nnoise= \Phi^{-1}(F( \Enoise ))$ into \eqref{equ:Trans} to get
\begin{align}
	\transfun{\Phi^{-1}(F( \Enoise ))} & =  F^{-1}(\Phi(\Phi^{-1}(F( \Enoise )))) =  \Enoise.
\end{align}

To get the desired approximation we use a \ac{PCHIP}  \cite{monotonecubic} to inter- and extrapolate. The \ac{PCHIP} function ($f_p(\cdot)$) is fitted so that
\begin{equation}
	\pchipfun{ \Phi^{-1}(\PE(\sample_i))} = \sample_i.
\end{equation}  
\acp{PCHIP} have the property that if the values are monotonic then the resulting function is monotonic and the first derivative is continuous. The \ac{PCHIP} algorithm in \cite{monotonecubic} fits the spline by interpolation at given knots. We propose to fit the \ac{PCHIP} model using integer standard deviation points as the knots, for example, $-1\sigma, 0, 1\sigma$. A \ac{PCHIP} is defined with the values and slopes of the interpolating function at these knots. In our application function $\pchipfun{\cdot}$ should be defined for all parameter values, and not only between the points that are used for data fitting. To achieve this we use linear extrapolation. 

The algorithm presented in~\cite{monotonecubic} has some open design parameters. Our implementation of choosing knots and computing slopes is given in Algorithm~\ref{algo:PCHIP} and the interpolation computation is given in Algorithm~\ref{algo:PCHIPval}.

\begin{algorithm}[!tb]
\caption{Slope computation for \ac{PCHIP}}
\label{algo:PCHIP}
	\SetKwInOut{Input}{input}
	\SetKwInOut{Output}{output}
	\Input{ $n$ -- number of samples\\
	           $\varepsilon_{E,j}$ -- $n$ measurement errors }
	\Output{ $s_i$ -- knot locations \\ $y_i$ -- values at knots \\ $d_i$ -- slopes at knots }
	Compute knot locations:
 	$\sample_1 = \lceil \Phi^{-1}(1/(n+1)) \rceil$\\
 	$m \leftarrow 1-2\sample_1$ \tcp{Number of Knots}
 	$\sample_i = \sample_1 + i - 1$ \\

	Select function values: 
	$\knotval_i \leftarrow \mathrm{quantile}(\varepsilon_E, \Phi{sample_i}) $ \\
	
	$k_j \leftarrow \Phi^{-1}\left(\frac{ | \{\varepsilon_{E,l} |  \varepsilon_{E,l} <  \varepsilon_{E,j} \}| }{n+1}\right)$ \tcp{Compute how many sigmas each sample is from the mean}

	\tcp{Initialize derivatives at knots by fitting a slope to samples in the interval}
	\For{$i=1$ to $\polypoints$} 
	{
	$\derivative_i \leftarrow \frac{\sum_{j,k_j > \sample_{i-1} \& k_j \leq \sample_{i+1} }{ (k_j - \sample_i)(\varepsilon_{E,j}-\knotval_i )}}{\sum_{j,k_j > \sample_{i-1} \& k_j \leq \sample_{i+1} }(k_j - \sample_i)^2}  $
	
	}
	\tcp{Check and alter derivatives if necessary}
	\For{$i=1$ to $\polypoints-1$} 
	{
	$\slope \leftarrow  \frac{\knotval_{i+1} - \knotval_{i}}{\knotpos_{i+1} - \knotpos_{i}}$ \\
	$\alpha \leftarrow  \frac{\derivative_i}{\slope}$ \\
	$\beta \leftarrow  \frac{\derivative_{i+1}}{\slope}$ \\
	\tcp{Alter derivatives if the spline would  not be monotonic}
	\If{$\sqrt{\alpha^2+\beta^2}>3$}
	{
	$\tau \leftarrow \frac{3}{\sqrt{\alpha^2+\beta^2}}$ \\
	$d_i \leftarrow \tau d_i$ \\
	$d_{i+1} \leftarrow \tau d_{i+1}$
	}
	}
\end{algorithm}

\begin{algorithm}[!tb]
\caption{Interpolation computation for \ac{PCHIP}}
	\SetKwInOut{Input}{input}
	\SetKwInOut{Output}{output}
	\Input{ $y_i$ -- values at knots \\ $d_i$ -- slopes at knots \\ $s_i$ -- knot locations \\ $s$ --  where to evaluate spline}	\Output{ $y$ -- value of spline at $s$  }
	
\label{algo:PCHIPval}
	\uIf {$\knotpos \leq \knotpos_1 $}
		{$\knotval \leftarrow \knotval_1 + \derivative_1(\knotpos-\knotpos_1)$}
	\uElseIf {$\knotpos \geq \knotpos_\polypoints $}
		{$\knotval \leftarrow \knotval_\polypoints + \derivative_\polypoints(\knotpos-\knotpos_\polypoints)$}
	\Else {
		Find $i$ so that $\knotpos_i < \knotpos \leq \knotpos_{i+1}$ \\
		$h_- \leftarrow \frac{\knotpos-\knotpos_{i}}{\knotpos_{i+1}-\knotpos_i}$ \\
$h_+ \leftarrow \frac{\knotpos_{i+1}-\knotpos}{\knotpos_{i+1}-\knotpos_i}$ \\
$\knotval \leftarrow \knotval_i (3h_+^2-2h_+^3) + \knotval_{i+1} (3h_-^2-2h_-^3)  - \derivative_i(\knotpos_{i+1}-\knotpos_i)(h_+^3-h_+^2)+\derivative_{i+1}(\knotpos_{i+1}-\knotpos_i)(h_-^3-h_-^2)$
	}	
\end{algorithm}
	
\begin{figure}
\includegraphics[width=\columnwidth]{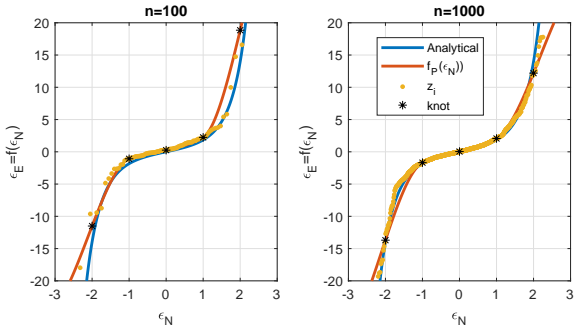}
\caption{A \ac{PCHIP} fitted to 100 and 1000 samples from a Student-$t$ distribution}
\label{fig:pchipfit}
\end{figure}

Figure~\ref{fig:pchipfit} shows an example of a fitted \ac{PCHIP}. It is fitted to 100 and 1000 samples from a Student-$t$ distribution with 1 degree of freedom, which means that it does not have mean and variance. Data points are shown with yellow dots and the knots with black asterisks. The blue curve shows the analytical transformation and the red curve the \ac{PCHIP} model. The figure shows how the curve fits the data better when the number of samples is increased.

\section{Kalman filtering with transformed noise}
\label{sec:ite}
Most Kalman filter extensions, such as \ac{EKF} \cite{nonlEKF}  or \ac{UKF}  \cite{WANUKF}, are suitable for estimation  with nonlinearly transformed noise. However, most of these extensions do a linearization at the prior mean. When we consider measurement functions of the form \eqref{equ:measnoise}, the $\transfun{\Nnoise}$ term has the same prior at every time instance and could be linearized offline as well and replaced with a normal approximation. By this way, the estimation result would not take the distribution of the empirical model into account.

We propose to use an iterative algorithm to obtain linearization that is not always in the same. This was done in \cite{GRUF} by extending of the \ac{RUF} \cite{RUKF,UKFRUF} to handle augmented states. However, in \cite{GRUF} it was found that in certain situations the \ac{RUF}-based filters did not give good results. Here we develop an algorithm for filtering with the proposed empirical error model based on \ac{IPLF} \cite{PLF}. The \ac{IPLF} algorithm computes iteratively linearizations based on the posterior estimates. 

When a posterior estimate (which on the first iteration is the prior) that has known mean $\statemean_i$ and covariance $\statecov_i$, the following expectations are approximated
\small
\begin{align}
	\predmeas &= \int \measfun(\state)p_{\N}(\state ; \statemean_i, \statecov_i) \d \state \label{equ:i1}\\
	\Pyx &= \int  \left( \measfun(\state)  - \predmeas\right)\left( \state - \statemean_i\right)^Tp_{\N}(\state ; \statemean_i, \statecov_i) \d \state \label{equ:i2} \\
	\Pyy &= \int \left( \measfun(\state)  - \predmeas\right)\left( \measfun(\state)  - \predmeas\right)^Tp_{\N}(\state ; \statemean_i, \statecov_i)  \d \state \label{equ:i3},
\end{align}
\normalsize
where $p_{\N}$ is the \ac{pdf} of the normal distribution.
These expectations are usually approximated using a sigma-point method that turns the integrals into finite sums.
In \ac{IPLF}, the expectations \eqref{equ:i1}-\eqref{equ:i3} are used to generate a linear measurement that corresponds to those moments
\cite{PLF}:
\begin{equation}
	\hat h(x) = \jacobian_i \state + \constant_i + \nonlnoise_i + \measnoise , \label{equ:meas2}
\end{equation}
where $\measnoise$ is the measurement noise and 
\begin{align}
\jacobian_i &= \Pxy_i^T \Pxx_i^{-1} \label{equ:jacobian}\\ 
\constant_i &=  \predmeas_i - \jacobian_i \statemean_{i}  \\
\nonlcov_i &=  \Pyy_i - \jacobian_i \Pxx_i \jacobian_i^T \label{equ:noncov} \\ 
\nonlnoise_i &\sim \N(0, \nonlcov_i),
\end{align}
where  $\nonlnoise_i$
is an additional noise independent from $\epsilon$, whose covariance $\nonlcov_i$ depends on the amount of the nonlinearity in the linearization region. For linear systems $\nonlcov_i$ is 0. The linearized measurement function \eqref{equ:meas2} is then used to update the prior and obtain the mean and covariance of the posterior as follows
\begin{align}
	\innocov &= \jacobian_i \priorcov \jacobian_i^T + \noisecov + \nonlcov_i \label{equ:innocov}\\ \
	\kalmangain_i & =  \priorcov \innocov^{-1}\\
	\statemean_{i+1} & = \priormean + \kalmangain_i(\meas - \jacobian_i\priormean - \constant_i) \label{equ:mean}\\
	\statecov_{i+1} & = \priorcov - \kalmangain_i\innocov_i\kalmangain_i^T, \label{equ:posteriorcov}
\end{align}
where $\priorcov$ is the prior covariance. The obtained posterior (\eqref{equ:mean} and \eqref{equ:posteriorcov}) is used for the next linearization. The process is repeated until a convergence criterion or maximum number of iterations is met.

\ac{IPLF}  is not directly suitable for our purposes as it assumes additive Gaussian noise. A standard approach for estimation with non-additive noise  is to use an augmented state. The augmented state $\augstate$ contains the state and error variables:
\begin{equation}
	\augstate = \matr{\state\\\noise_{\N}}\sim \N \left( \matr{\statemean \\ \noisemean}, \matr{ \Pxx & \Pxe \\\Pex & \Pee} \right).
\end{equation}
However, using an augmented state iteratively in \ac{IPLF} may cause problems. As the noise variables are also in the state, it is possible that the covariance of the augmented state $\Pzz$ computed using \eqref{equ:posteriorcov} is singular. This happens if measurements are linear, but may happen also due to inaccurate approximation of the integrals \eqref{equ:i2}-\eqref{equ:i3}. Approximation of integrals \eqref{equ:i1}-\eqref{equ:i3} requires extra care if the covariance is singular or close to singular. However bigger problems occur in computation of the Jacobian in \eqref{equ:jacobian} as it contains an inverse of a singular matrix. However, even if the covariance matrix of the augmented state is singular the part of it that corresponds to the state variables is not and the posterior after the iterations is not singular (unless there are noise-free measurements). We propose to avoid the singularity problem by adding a diagonal matrix with small positive elements to the augmented covariance. The small positive elements can be chosen to be fractions of the diagonal 
\begin{equation}
	{\hatPzz}_i = \Pzz_i + \kappa\,\diag{\Pzz_i}, \label{equ:roundcov}
\end{equation}
where $\kappa>0$. We use $\hatPzz_i$ in equations corresponding to \eqref{equ:i1}-\eqref{equ:i3}  and \eqref{equ:jacobian}, thus the larger covariance affects the linearization only by taking the function values from a larger area into account. If the measurement is linear, the linearization is exact, independent of $\kappa$. Another option for dealing with singular covariance matrices would be  to add small constants to eigenvalues that are zero (or negative due to roundoff errors), but this method would be computationally more demanding. We will show that \eqref{equ:roundcov} works well with \ac{PCHIP} noise models.

\begin{figure}
\includegraphics[width=\columnwidth,clip=true,trim=0cm 0cm 0cm 0cm]{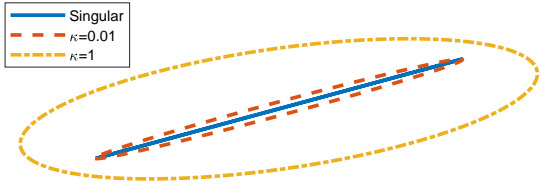}
\caption{How value of $\kappa$ affects $1\sigma$ ellipsoids when the original distribution is singular}
\label{fig:alphaeffect}
\end{figure}

Figure~\ref{fig:alphaeffect} show the effect of $\kappa$ values 0.01 and 1 to  $1\sigma$ ellipsoids. The original covariance is singular and, thus, the ellipsoid is collapsed to a line. When the $\kappa$ value is increased the ellipse becomes wider.

When fitting the \ac{PCHIP} model most of the samples are located near the mean. To have a sample further than $5\sigma$ one needs about $10^6$ samples, and to have a data point further than $6\sigma$ the required number of points is more than $10^9$. The \ac{PCHIP} model is not accurate far from the mean because there was little or no data there when building the model. This makes use of certain \acp{CKF} impractical as the sigma-point spread depends on the number of state dimension $\statedim$ and measurement dimension $\measdim$ as $\sqrt{\statedim + \measdim}$, because the non-Gaussian measurements are used in the augmented state \cite{cubature}. Due to this we use \ac{UKF} with a small sigma-point spread.

The noise dimensions of the augmented state have always unit covariance in the first iteration of the filter. With heavy-tailed distributions, for example with Student-$t$ shown in Figure~\ref{fig:pchipfit} the estimate could sometimes end up during iterations in regions where there were not statistically enough samples.

To address the problem of getting mean outside regions of where samples exist in iterations, we propose to use a simplified version of \ac{DPLF}~\cite{dampedPLF}. In \ac{DPLF} the value of a cost function is optimized by a line search algorithm that makes the state mean change less in the update if necessary. The problem with \ac{PCHIP} is similar, but instead optimizing a cost function we want to ensure that the estimate does not jump into an area where samples are sparse. In update of the mean in \cite{dampedPLF} there is $0<\alpha\leq1$
that governs how much the mean changes in one iteration:
\begin{equation}
\statemean_{i+1} = (1-\alpha)\statemean_{i}+ \alpha(\statemean_0 + K_{i}( y - \measfun(\statemean_{i} ))). \label{equ:meanup}
\end{equation}
When  when $alpha$ is 1 this corresponds to \eqref{equ:mean}. Recalling that $\Nnoise$ has always prior variance 1, we propose to choose $\alpha$ so that the mean corresponding to the noise variables never changes more than 1 (which corresponds to a change of $1\cdot\sigma$).

The full algorithm for iterative estimation with augmented state is presented in Algorithm~\ref{algo:iplf2}. 

\begin{algorithm}
\caption{Iterative update of the state with transformed measurement noise}
\label{algo:iplf2}
	$i\leftarrow1$\\
	$\augmean_i \leftarrow \matr{\statemean^- \\ 0}$ \\
	 $\Pzz_i \leftarrow \matr { \Pxx & 0 \\ 0 & I} $\\
	\While{$i \leq \iterations$}
	{
	$\hatPzz_i \leftarrow \Pzz_i + \kappa \diag{\Pzz_i}$ \\  
	Compute approximations to integrals using \ac{UKF}: \\ \Indp
	{
	$\predmeas_i \leftarrow \int \measfun(\augstate)p_{\N}(\augstate ; \augmean_i, \hatPzz_i) \d \augstate$\\
	$\Pyz_i \leftarrow \int  \left( \measfun(\augstate)  - \predmeas_i\right)\left( \augstate - \augmean_i\right)^Tp_{\N}(\augstate ; \augmean_i, \hatPzz_i) \d \augstate$ \\
	 $\Pyy \leftarrow \int \left( \measfun(\augstate)  - \predmeas_i\right)\left( \measfun(\augstate)  - \predmeas_i\right)^Tp_{\N}(\augstate ; \augmean_i, \hatPzz_i)  \d \augstate $\\
	 } \Indm
	 Compute parameters for linearized measurement: \\\Indp
	 {
	 	$\jacobian_i\leftarrow \Pzy_i^T {\hatPzz}_i^{-1}$ \\
		$\constant_i\leftarrow \predmeas_i  - \jacobian_i \augmean_{i}$  \\
		$\nonlcov_i \leftarrow \Pyy_i - \jacobian_i{\hatPzz}_i\jacobian_i^T$ \\
	 }\Indm
	 Compute posterior estimate for the next iteration: \\\Indp
	 {
	 	$\innocov_i \leftarrow \jacobian_i \Pzz_0 \jacobian_i^T + \nonlcov_i$ \\
		$\gain_i \leftarrow \augcov_0 \jacobian_i^T \innocov_i^{-1}$ \\
		$\Delta\augmean_i =\gain_i( \meas - \jacobian_i \augmean_0 - \constant_i)$ \\
		$ \alpha = \min \left(\frac{1}{\max \Delta\augmean_{i,\Nnoise}}, 1\right)$ \tcp{where $\max \Delta\augmean_{i,\Nnoise}$ is the maximum change of the noise variables in the augmented state } 
		$\augmean_{i+1} \leftarrow (1-\alpha)\augmean_i+ \alpha( \augmean_0 + \Delta\augmean_i) $ \\
		$\augcov_{i+1} \leftarrow \augcov_0 - \gain_i\innocov_i\gain_i^T$\\
	 }
		$i\leftarrow i+1$	 
	}
	 Extract posterior of the state from the augmented state\\
	 $\statemean^+  \leftarrow \matr{I_{\statedim \times \statedim} & \mathbf{0}_{\statedim \times \measdim}}\augmean_i$ \\
	 $\statecov^+  \leftarrow \matr{I_{\statedim \times \statedim} & \mathbf{0}_{\statedim \times \measdim}} \Pzz \matr{I_{\statedim \times \statedim} & \mathbf{0}_{\statedim \times \measdim}}^T$ \\

\end{algorithm}

\section{Results}
\label{sec:results}
\subsection{Simulated Student-$t$ test}

In our first test, we consider the fourth example from \cite{GRUF}. The state is two-dimensional with state transition model 
\begin{equation}
	x_{t+1} =  \matr{1&1\\0&1 } x_t + \varepsilon_Q,
\end{equation}
where $\varepsilon_Q \sim \N\left(   \bigl[ \begin{smallmatrix} 0\\0 \end{smallmatrix} \bigr] , \bigl[ \begin{smallmatrix} 0&0\\0&1 \end{smallmatrix} \bigr]   \right)$ and $x_0 \sim \N\left(  \bigl[ \begin{smallmatrix} 0\\0 \end{smallmatrix} \bigr] ,  \bigl[ \begin{smallmatrix} 40&0\\0&4  \end{smallmatrix} \bigr]  \right)$.
The measurements are noisy observations of the first state variable
\begin{equation}
	y_t =\matr{1 & 0}x_t+\Enoise,
\end{equation}
where $\Enoise$ is  Student-$t$ distributed with 3 degrees of freedom scaled with $\sqrt{\frac{100}{3}}$. The variance of $\Enoise$ is thus 100. We simulated 100 sequences, each 50 time steps long.

In the second test we test how the number of samples used for generating the function $\pchipfun\cdot$ and the number of iterations in Algorithm~\ref{algo:iplf2} change the accuracy of the estimation. 
\begin{figure}
	\includegraphics[width=\columnwidth]{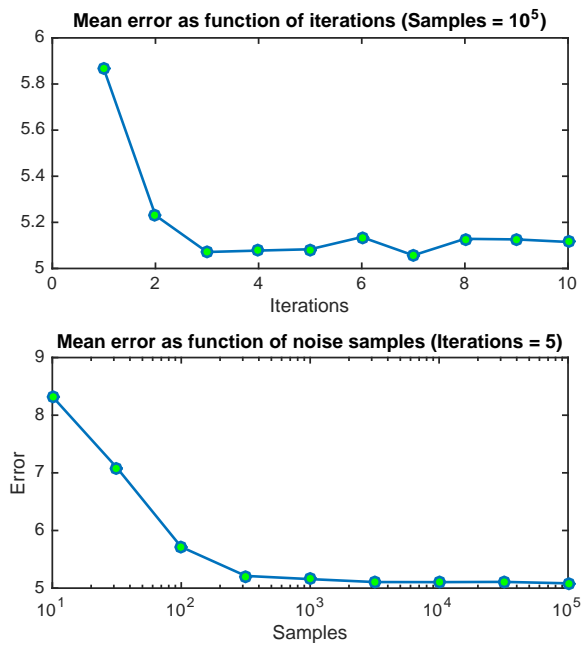}
	\caption{Mean of absolute errors as function of iterations of Algorithm~\ref{algo:iplf2} and as function of the count of samples used to fit \ac{PCHIP}}
	\label{fig:errorasfunction}
\end{figure}
The estimation accuracy  does not further improve after using more than 3 iterations. For testing  $\pchipfun\cdot$ we generated 10 random sets of samples for fitting the \ac{PCHIP} model. Also, there is no major improvement after having 300 samples of the error distribution.

Table~\ref{tbl:err} shows the mean errors of the proposed algorithms with $10^3$ and $10^5$ samples using 5 iterations and averages of absolute errors obtained using \ac{KF} that uses Gaussian noise with same mean and variance as the Student-$t$ noise (mean 0 and variance 100), \ac{RORF} \cite{piche2012d}, and \ac{GRUF} \cite{GRUF}. 
\begin{table}
\caption{Mean errors of state varibables in Student-$t$ test}
\label{tbl:err}
\begin{tabular}{c|ccccc}
		    & \ac{KF} & \ac{GRUF} & \ac{RORF} & Proposed & Proposed  \\
	&	    & &  &$\nnoise=10^3$&$\nnoise=10^5$\\ \hline
    Mean error $x_{[1]}$ &6.50 & 5.33 & 5.02 & 5.16 & 5.12    \\
Mean error $x_{[2]}$ & 2.74 & 2.08 & 2.03 & 2.05 & 2.04 
\end{tabular}
\end{table}
We use only 2 iterations in \ac{GRUF}  as we found that the error is then the smallest. When comparing to Figure~\ref{fig:errorasfunction} we can see that the proposed algorithm outperforms \ac{GRUF} also with 2 iterations. Table~\ref{tbl:err} shows that only \ac{RORF} has better accuracy than the proposed algorithm.  However, the marginal between \ac{RORF} and the proposed algorithm is only one third of the marginal between \ac{GRUF} and \ac{RORF} and \ac{RORF} is designed specifically for Student-$t$ noise. The proposed algorithm is computationally most demanding of the tested algorithms, but it is of same order as other iterative algorithms.

\subsection{Low-Earth orbiting satellite magnetometer measurements}

\begin{figure}
	\includegraphics[width=\columnwidth]{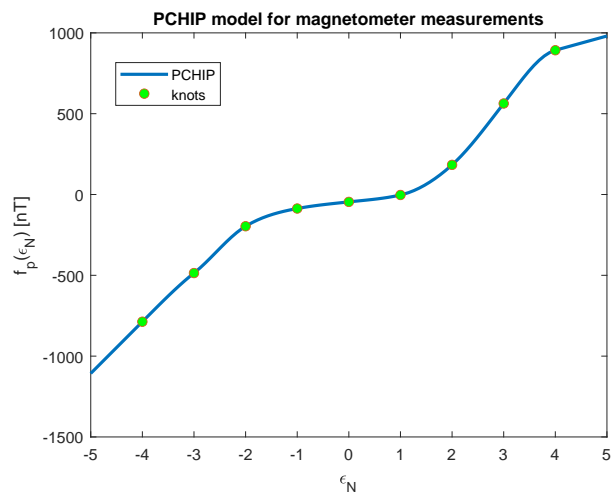}
	\caption{\ac{PCHIP} model for residuals of magnetometer measurements}
	\label{fig:magnetometer}
\end{figure}

In this section, we build \ac{PCHIP} models for real data collected from satellite magnetometers. The satellite magnetometer measurements from SWARM-A satellite \cite{demet} are compared with \ac{IGRF} model and we fit the proposed model to the residuals. Figure~\ref{fig:magnetometer} shows the \ac{PCHIP} model fitted to these residuals.
We can see from the figure that the slope of the \ac{PCHIP} model is smallest when $\varepsilon_N$ is close to 0, which means that the distribution of the residuals has heavier tails than the normal distribution.

For a test with simulated measurements, we assume a satellite that has magnetometer and sun sensor. Both sensors are modeled using 
\begin{equation}
	y = A(x)z_\text{model} + \varepsilon,
\end{equation}
where $z_\text{model}$ is a vector that is obtained from a magnetic field model or a model where the sun should be at the current location. In this paper, we omit the position part in the modeling. Furthermore, we omit the normalization of the model vectors. In reality, the sun vector is a direction without a magnitude. 

Magnetometers measure the magnetic field vector whereas, sun sensors measure the current output, which is processed to give the direction of the Sun. In this study, magnetometer model values and the measurement noises are taken from real measurements, but randomized, while the sun sensor measuremenets is modeled using random normal distributed measurements and the random noise associated with it is also a normal distributed. The state is three dimensional and contains pitch $\pitch$, yaw $\yaw$, and roll $\roll$. The matrix $A(x)$ that transforms the model vectors from reference to body coordinates is
\begin{align}& A(x)  =  \\&  \smatr{\cos\yaw\cos\pitch & \cos\yaw\sin\pitch & -\sin\yaw\\
   -\cos\roll\sin\pitch+\sin\roll\sin\yaw\cos\pitch & \cos\roll\cos\pitch+\sin\roll\sin\yaw\sin\pitch & \sin\roll\cos\yaw\\
   \sin\roll\sin\pitch+\cos\roll\sin\yaw\cos\pitch & -\sin\roll\cos\pitch+\cos\roll\sin\yaw\sin\pitch & \cos\roll\cos\yaw} \notag
\end{align}
and the purpose of the estimation algorithm is to get estimates for $\pitch$, $\yaw$, and $\roll$.

\begin{figure}
	\includegraphics[width=\columnwidth,clip=true,trim=1.5cm 0cm 1.9cm 0cm]{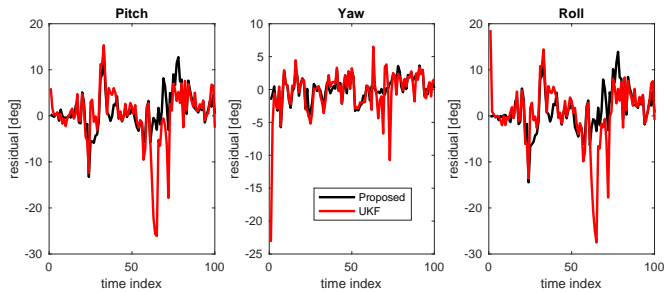}
	\caption{Residuals of estimates of pitch, yaw, and roll in one simulated test}
	\label{fig:satellite}
\end{figure}

In our test, we simulate the Euler angles of a tumbling satellite as pitch $\pitch$, yaw $\yaw$, and roll $\roll$. Figure~\ref{fig:satellite} shows an example estimation errors of the desired variables with the proposed algorithm and with \ac{UKF}.
The fast convergence of the state at the time index is due to the iterative nature of the estimation algorithm as the first prior is poor and the linearization is also poor. At around time index 60 the proposed algorithm also has smaller residual than \ac{UKF}, which is caused by the ability of the proposed algorithm to cope with the outliers in the data.

\subsection{\ac{UWB} measurements}

 The \ac{UWB} system measures the distance between a transmitter at known position $\transloc$ and receiver $\recloc$
\begin{equation}
  \measfun(\recloc,\Enoise) = \| \transloc - \recloc \| + \Enoise . \label{equ:UWB}
\end{equation}
The noise $\Enoise$ in \ac{UWB} measurements is not normal and its error characteristics vary depending on signal conditions. Measurements tend to have positive bias and large variance in \ac{NLOS} conditions \cite{4529090}. 

In~\cite{Kok2015} the measurement error is modeled using an asymmetric \ac{pdf}. The positioning problem is then solved using an iterative Gauss-Newtown algorithm for the whole track. In~\cite{henriletter}  \ac{UWB} ranging errors are modeled with a skew-$t$ distribution. The algorithm in~\cite{henriletter} is a filtering algorithm, thus, the state estimate can be computed online when new measurements arrive. 

\begin{figure}
\includegraphics[width=\columnwidth]{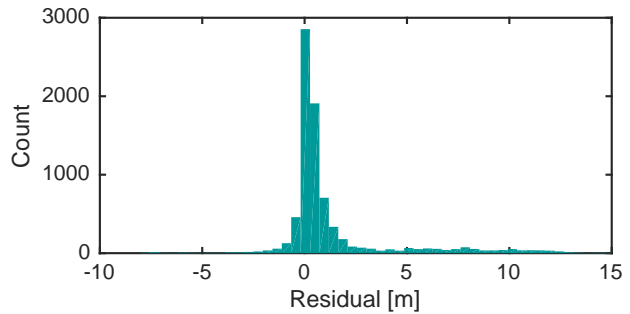}
\caption{Residuals of \ac{UWB} measurements}
\label{fig:residuals}
\end{figure}

In this work, we use the same measurements data set as in~\cite{henriletter}. Measurements are divided into learning and verification sets. Learning set contains 6 separate tracks and verification set 3 tracks. The tracks were collected by walking with a \ac{UWB} receiver in hand. The reference positions were collected using a Vicon real-time tracking system. The distribution of noise is determined using the learning data
\begin{equation}
\varepsilon_{\mathrm{E},i} =  y_i - \| \transloc - \recloc \|,
\end{equation} 
where $y_i$ is the measurement value obtained at location $\recloc$. Figure~\ref{fig:residuals} shows the histogram of \ac{UWB} measurement residuals $\Enoise$ of the learning set.

\begin{figure}
\centering 
\includegraphics[width=0.9\columnwidth,clip=true,trim=1cm 1.5cm 1cm 0.5cm]{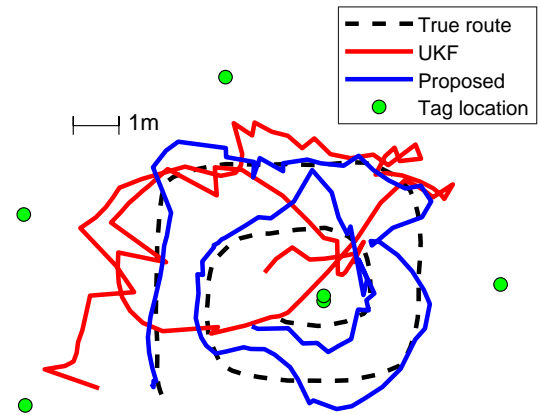}
\caption{An example route}
\label{fig:pedestrianroute}
\end{figure}

The proposed model learned from \ac{UWB} residuals is used for pedestrian positioning. Test setup uses same data as  \cite{henriipin}, except that we do not use inertial measurements. Fig.~\ref{fig:pedestrianroute} shows a part of one of our test routes computed with the proposed error model using \ac{IPLF} and with \ac{UKF} with Gaussian noise assumption.
On the tested tracks the proposed algorithm had an average error of $0.5$\,m, while estimate that used Gaussian assumption had an average error of $1.5$\,m. This shows how the proposed algorithm is clearly more accurate than a filter that usees Gaussian assumption. The errors of the proposed algorithm are of the same order as the results obtained in \cite{henriipin}, where also  inertial sensors were used.

\section{Conclusions}
\label{sec:conclusions}
Most Kalman filter extensions assume normal distributed noise. We proposed a method for transforming empirically measured residuals of a non-Gaussian noises into a function that can be used in estimation with Kalman filter extensions. The proposed method fits noise samples using \acp{PCHIP} and can be used with various types of distributions.

Our simulation and real-data results showed that the proposed algorithm improved estimation accuracy compared to algorithms that assumed Gaussian noise, and had similar estimation accuracy as algorithms that were developed specifically for certain noise distributions.

\section{Acknowledgements}
High accuracy reference measurements are provided through the use of the Vicon real-time tracking system courtesy of the UAS Technologies Lab, Artificial Intelligence and Integrated Computer Systems Division (AIICS) at the Department of Computer and Information Science (IDA). \url{http://www.ida.liu.se/divisions/aiics/aiicssite/index.en.shtml}

The magnetometer data was obtained from \url{ftp://swarm-diss.eo.esa.int/}.

\balance
\bibliographystyle{IEEEtran}
\bibliography{viitteetPLF}

\begin{thebibliography}{10}
\providecommand{\url}[1]{#1}
\csname url@samestyle\endcsname
\providecommand{\newblock}{\relax}
\providecommand{\bibinfo}[2]{#2}
\providecommand{\BIBentrySTDinterwordspacing}{\spaceskip=0pt\relax}
\providecommand{\BIBentryALTinterwordstretchfactor}{4}
\providecommand{\BIBentryALTinterwordspacing}{\spaceskip=\fontdimen2\font plus
\BIBentryALTinterwordstretchfactor\fontdimen3\font minus
  \fontdimen4\font\relax}
\providecommand{\BIBforeignlanguage}[2]{{%
\expandafter\ifx\csname l@#1\endcsname\relax
\typeout{** WARNING: IEEEtran.bst: No hyphenation pattern has been}%
\typeout{** loaded for the language `#1'. Using the pattern for}%
\typeout{** the default language instead.}%
\else
\language=\csname l@#1\endcsname
\fi
#2}}
\providecommand{\BIBdecl}{\relax}
\BIBdecl

\bibitem{outliers}
R.~Pearson, ``Outliers in process modeling and identification,'' \emph{Control
  Systems Technology, IEEE Transactions on}, vol.~10, no.~1, pp. 55--63, Jan
  2002.

\bibitem{henriletter}
H.~Nurminen, T.~Ardeshiri, R.~Piche, and F.~Gustafsson, ``Robust inference for
  state-space models with skewed measurement noise,'' \emph{Signal Processing
  Letters, IEEE}, vol.~22, no.~11, pp. 1898--1902, Nov 2015.

\bibitem{2015arXiv150904072W}
M.~W{\"u}thrich, C.~Garcia~Cifuentes, S.~Trimpe, F.~Meier, J.~Bohg, J.~Issac,
  and S.~Schaal, ``Robust {G}aussian filtering,'' in \emph{Proceedings of the
  American Control Conference}, Boston, MA, USA, Jul. 2016.

\bibitem{agamennoni}
G.~Agamennoni, J.~Nieto, and E.~Nebot, ``An outlier-robust {K}alman filter,''
  in \emph{IEEE International Conference on Robotics and Automation (ICRA)},
  May 2011, pp. 1551--1558, \doi{10.1109/ICRA.2011.5979605}.

\bibitem{piche2012d}
R.~Pich{\'e}, S.~S{\"a}rkk{\"a}, and J.~Hartikainen, ``Recursive outlier-robust
  filtering and smoothing for nonlinear systems using the multivariate
  {S}tudent-t distribution,'' in \emph{IEEE International Workshop on Machine
  Learning for Signal Processing (MLSP)}, 2012, pp. 1--6,
  \doi{10.1109/MLSP.2012.6349794}.

\bibitem{GRUF}
M.~Raitoharju, R.~Pich\'e, and H.~Nurminen, ``A systematic approach for
  kalman-type filtering with non-gaussian noises,'' in \emph{2016 19th
  International Conference on Information Fusion (FUSION)}, July 2016, pp.
  1853--1858.

\bibitem{8260875}
F.~{Tronarp}, .~F. {García-Fernández}, and S.~{Särkkä}, ``Iterative
  filtering and smoothing in nonlinear and non-gaussian systems using
  conditional moments,'' \emph{IEEE Signal Processing Letters}, vol.~25, no.~3,
  pp. 408--412, 2018.

\bibitem{RAITOHARJU2020107330}
M.~Raitoharju, A.~Garc\'ia-Fern\'andez, R.~Hostettler, R.~Pich\'e, and
  S.~S\"arkk\"a, ``Gaussian mixture models for signal mapping and
  positioning,'' \emph{Signal Processing}, vol. 168, p. 107330, 2020.

\bibitem{monotonecubic}
F.~N. Fritsch and R.~E. Carlson, ``Monotone piecewise cubic interpolation,''
  \emph{SIAM Journal on Numerical Analysis}, vol.~17, no.~2, pp. 238--246,
  1980.

\bibitem{nonlEKF}
D.~Simon, \emph{Optimal State Estimation: Kalman, H Infinity, and Nonlinear
  Approaches}.\hskip 1em plus 0.5em minus 0.4em\relax Wiley-Interscience, 2006.

\bibitem{WANUKF}
E.~Wan and R.~Van~der Merwe, ``The unscented {K}alman filter for nonlinear
  estimation,'' in \emph{Proc. Adaptive Systems for Signal Process., Commun.,
  and Control Symp.. AS-SPCC.}, 2000, pp. 153--158,
  \doi{10.1109/ASSPCC.2000.882463}.

\bibitem{RUKF}
R.~Zanetti, ``Recursive update filtering for nonlinear estimation,'' \emph{IEEE
  Trans. Autom. Control}, vol.~57, no.~6, pp. 1481--1490, June 2012,
  \doi{10.1109/TAC.2011.2178334}.

\bibitem{UKFRUF}
Y.~Huang, Y.~Zhang, N.~Li, and L.~Zhao, ``\BIBforeignlanguage{English}{Design
  of sigma-point {K}alman filter with recursive updated measurement},''
  \emph{\BIBforeignlanguage{English}{Circuits, Systems, and Signal Process.}},
  pp. 1--16, August 2015, \doi{10.1007/s00034-015-0137-y}.

\bibitem{PLF}
{\'A.F}.~Garc\'ia-Fern\'andez, L.~Svensson, M.~Morelande, and S.~S\"arkka,
  ``Posterior linearization filter: Principles and implementation using sigma
  points,'' \emph{IEEE Transactions on Signal Processing}, vol.~63, no.~20, pp.
  5561--5573, Oct 2015, \doi{10.1109/TSP.2015.2454485}.

\bibitem{cubature}
I.~Arasaratnam and S.~Haykin, ``Cubature {K}alman filters,'' \emph{IEEE Trans.
  Autom. Control}, vol.~54, no.~6, pp. 1254--1269, June 2009,
  \doi{10.1109/TAC.2009.2019800}.

\bibitem{dampedPLF}
M.~{Raitoharju}, L.~{Svensson}, A.~F. {Garc\'ia-Fern\'andez}, and R.~{Pich\'e},
  ``Damped posterior linearization filter,'' \emph{IEEE Signal Processing
  Letters}, vol.~25, no.~4, pp. 536--540, April 2018.

\bibitem{demet}
D.~Cilden-Guler, M.~Raitoharju, R.~Piche, and C.~Hajiyev, ``Nanosatellite
  attitude estimation using {K}alman-type filters with non-{G}aussian noise,''
  \emph{Aerospace Science and Technology}, vol.~92, pp. 66 -- 76, 2019.

\bibitem{4529090}
N.~Alsindi, B.~Alavi, and K.~Pahlavan, ``Measurement and modeling of
  ultrawideband toa-based ranging in indoor multipath environments,''
  \emph{Vehicular Technology, IEEE Transactions on}, vol.~58, no.~3, pp.
  1046--1058, March 2009.

\bibitem{Kok2015}
M.~Kok, J.~D. Hol, and T.~B. Sch\"{o}n, ``Indoor positioning using
  ultrawideband and inertial measurements,'' \emph{IEEE Transactions on
  Vehicular Technology}, vol.~64, no.~4, pp. 1293--1303, April 2015.

\bibitem{henriipin}
H.~Nurminen, T.~Ardeshiri, R.~Pich{\'e}, and F.~Gustafsson, ``A nlos-robust toa
  positioning filter based on a skew-t measurement noise model,'' in \emph{2015
  International Conference on Indoor Positioning and Indoor Navigation (IPIN)},
  no.~11, November 2015, pp. 1898--1902.

\end{thebibliography}

\end{document}